\documentclass[dvipdfmx]{article}

\usepackage{amssymb,amsfonts,amsmath,amsthm,stmaryrd}
\usepackage{cite,enumerate,float,indentfirst}
\usepackage[dvipdfmx]{graphicx}
\usepackage{tikz}
\usepackage{color}
\usepackage{bm}
\usepackage{simplewick}
\usepackage{here}

\catcode`\@=11
\@addtoreset{equation}{section}

\theoremstyle{definition}

\theoremstyle{remark}

\allowdisplaybreaks

\newcommand{\beq}{\begin{equation}}
\newcommand{\eeq}{\end{equation}}
\newcommand{\bea}{\begin{eqnarray}}
\newcommand{\eea}{\end{eqnarray}}
\newcommand{\mat}[1]{\begin{pmatrix} #1 \end{pmatrix}}

\definecolor{red}{rgb}{1,0,0}
\definecolor{orange}{rgb}{1,0.5,0}
\definecolor{violet}{rgb}{0.7,0,1}

\hoffset= - 1.0in         % switch off for draft style
\begin{document}

\begin{center}
\begin{small}
\hfill NITEP 145
\end{small}
\end{center}

\hfill December, 2022

\bigskip

\begin{center}{\Large{\bf  A-D hypersurface of $su(n)$ $\mathcal{N}=2$ supersymmetric gauge theory\\
with $N_f = 2n-2$ flavors}
}
\end{center}

\bigskip

\centerline{\large H. Itoyama$^{a, b,}$\footnote{e-mail: itoyama@omu.ac.jp}, 
T. Oota$^{a,b,}$\footnote{e-mail: toota@omu.ac.jp}, and 
R. Yoshioka$^{a,b,}$\footnote{e-mail: ryoshioka@omu.ac.jp} }

\vspace{1cm}

\noindent
{$^a$ \it Nambu Yoichiro Institute of Theoretical and Experimental Physics (NITEP),
			Osaka Metropolitan University (formerly Osaka City University)}\\
{$^b$ \it Osaka Central Advanced Mathematical Institute (OCAMI), \\
\phantom{aaaaaaaa} 3-3-138, Sugimoto, Sumiyoshi-ku, Osaka, 558-8585, Japan}

\vspace{1cm}

\centerline{ABSTRACT}

\bigskip

%{\footnotesize
%}
In the previous letter, arXiv:2210.16738[hep-th], 
we found a set of flavor mass relations as constraints that the $\beta$-deformed $A_{n-1}$ quiver matrix model restores the maximal symmetry in the massive scaling limit
 and reported the existence of Argyres-Douglas critical hypersurface.
In this letter, we derive the concrete conditions on moduli parameters which maximally degenerates the Seiberg-Witten curve while maintaining the flavor mass relations. 
These conditions define the A-D hypersurface.

\bigskip

\bigskip

%%%%%%%%%%%%%%%%%%%%%%%%%%%%%%%%%%%%%%%%%%%%%%%%%%%%％ 
\section{Introduction}
%%%%%%%%%%%%%%%%%%%%%%%%%%%%%%%%%%%%%%%%%%%%%%%%%%%%％ 
The Seiberg-Witten system \cite{SW9407,SW9408} and matrix models have some interesting interplay in the development of supersymmetric gauge theory through integrability\cite{integrability}.  
In the series of the developments \cite{IOYone1008,IOYanok1805,IOYanok1812,IOYanok1909,IOYanok2019,IYanok2103} on the irregular limit 
\cite{Gaiotto0908,GT1203,MMM0909,BMT1112,NR1112,NR1207,CR1312,BGKNT2202} of matrix models with milti-log type potential, 
we have established  the ``unitarization" procedure 
which converts the hermitian matrix model into the unitary matrix model. 
In the previous letter \cite{IOY2210}, we considered the massive scaling limits of the $\beta$-deformed $A_{n-1}$ quiver matrix model and obtained the irregular conformal/W block which corresponds to $su(n)$ $\mathcal{N} = 2$ supersymmetric gauge theory with $N_f = 2n-2$ flavors \cite{AGT0906,Wyllard0907,MM0908}.  
After this limiting process,  
the potential contains the terms that break the maximal symmetry of the matrix model based on the automorphism of the $A_{n-1}$ Dynkin diagram. 
In order to restore the symmetry, 
 we imposed constraints on the parameters included in the potentials. 
We have identified the following flavor mass relations required  for maximal symmetry of the matrix model: 
\begin{equation} \label{mass}
  \tilde{m}_a = \tilde{m}_{2n-1-a}, ~~~ a= 1,2,\cdots, n-1, 
\end{equation}
where $\tilde{m}_a$ is the mass parameter of flavors.

Since the matrix model curve of this model and the associated Witten-Gaiotto curve are known to be isomorphic \cite{IMO}, the same conclusion should be drawn at the level of the S-W curve.
In keeping with this line, the purpose of the present letter is to provide 
 the conditions on the moduli parameters, under which the S-W curve is maximally degenerate in the region satisfying \eqref{mass}. 
These conditions define a hypersurface in the parameter space, which must be the one we called the A-D critical hypersurface in \cite{IOY2210}. 

This letter is organized as follows: 
In the next section, 
we introduce the S-W curve for $su(n)$, $N_f=2n-2$ susy gauge theory and its discriminant.
In section 3, we present the conditions that the moduli parameters have to satisfy in order for the S-W curve to be maximally degenerate. 
 
%%%%%%%%%%%%%%%%%%%%%%%%%%%%%%%%%%%%%%%%%%%%%%%%%%%%％ 
\section{S-W curve for $su(n)$, $N_f = 2n-2$ and its discriminant}
%%%%%%%%%%%%%%%%%%%%%%%%%%%%%%%%%%%%%%%%%%%%%%%%%%%%％ 
We consider the $su(n)$, $N_f=2n-2$ gauge theory with the mass relations \eqref{mass}. 
The Seiberg-Witten curve is given by  \cite{HO9505}
\begin{equation}\label{SWcurve}
 y^2 = C(\tilde{x})^2 - G(\tilde{x})^2,
\end{equation}
where 
\begin{align}
 C(\tilde{x}) = \tilde{x}^n + \sum_{i=2}^n \tilde{s}_i \tilde{x}^{n-i},  \label{C}\\
 G(\tilde{x}) = \Lambda \prod_{a=1}^{n-1} ( \tilde{x} + \tilde{m}_a). \label{G}
\end{align}
Here $\tilde{s}_i$ is the moduli parameter and $\Lambda$ is the scale parameter. 

Suppose that the S-W curve can be rewritten as 
\begin{equation}
 y^2 = P_-(\tilde{x}) P_+(\tilde{x}),
\end{equation}
where
\begin{equation}
 P_{\pm}(\tilde{x}) = \prod_{i=1}^n(\tilde{x} - f_i^{\pm}).
\end{equation}
We found that 
 the resultant of $P_-(\tilde{x})$ and $P_+(\tilde{x})$ with respect to the variables $\tilde{x}$ is given as 
\begin{equation}\label{resultant}
 R_{\tilde{x}}(P_-,P_+) 
  = \prod_{i=1}^n \prod_{j=1}^n (f_i^- - f_j^+) = (2\Lambda)^n \prod_{a=1}^{n-1} C(-\tilde{m}_a).
\end{equation} 
We have checked this relation up to $n=7$. 
The discriminant  of $P_-(\tilde{x})P_+(\tilde{x})$ with respect to $\tilde{x}$ is given by 
\begin{equation}\label{discrim}
\Delta_{\tilde{x}}(P_-P_+) = \Delta_{\tilde{x}}(P_-) \Delta_{\tilde{x}}(P_+) R_{\tilde{x}}(P_-,P_+)^2, 
\end{equation}
where $\Delta_{\tilde{x}}(P_{\pm})$ is the discriminant of $P_{\pm}(\tilde{x})$ with respect to $\tilde{x}$,
\begin{equation}
 \Delta_{\tilde{x}} (P_{\pm}) = \prod_{1\leq i<j\leq n} (f_i^{\pm} - f_j^{\pm})^2. 
\end{equation}

%%%%%%%%%%%%%%%%%%%%%%%%%%%%%%%%%%%%%%%%%%%%%%%%%%%%％ 
\section{A-D critical  hypersurface}
%%%%%%%%%%%%%%%%%%%%%%%%%%%%%%%%%%%%%%%%%%%%%%%%%%%%％ 
In this section, we derive the conditions that maximally degenerate the S-W curve \eqref{SWcurve}
 in the region satisfying \eqref{mass} for general $n$.  
 
In the case of $n=2$, the curve is given by 
\begin{equation}\label{curve:n=2}
 y^2 = (\tilde{x}^2 + {s})^2 - \Lambda^2 (\tilde{x} + m)^2, 
\end{equation}
where 
\begin{equation}
 {s} = -u + \frac{\Lambda^2}{8},~~~~~
 m = \tilde{m}_1 = \tilde{m}_2. 
\end{equation}
Here $u$ is  the Coulomb moduli parameter. 
The mass relation $\tilde{m}_1 = \tilde{m}_2$ is the same as taking all masses equal  
(see for example \cite{EHIY9603}).
Let $x = \tilde{x} + m$. Then eq.\eqref{curve:n=2} becomes
\begin{equation}
 y^2 = (x^2 + s_1 x + s_2)^2 - \Lambda^2 x^2,
\end{equation}
where 
\begin{equation}
\begin{split}
 &s_1 = -2m , \\
 &s_2 = s + m^2.
\end{split} 
\end{equation}
Imposing the following conditions on the parameters: 
\begin{align}
 &S_{21\pm} := s_1  \pm \Lambda = 0, \label{n=2:s_1} \\
 &S_{22} := s_2  = 0, \label{n=2:s_2}
\end{align}
we obtain 
\begin{equation}
 y^2 = x^3(x \mp 2 \Lambda).
\end{equation}
Further simplification by adjusting the parameters is impossible 
 and the curve is degenerate with multiplicity three. 
The conditions \eqref{n=2:s_1} and \eqref{n=2:s_2} should, therefore, provide A-D point \cite{AD9505,APSW9511,KY9712,Xie1204} in this case.
Note that the discriminant with respect to $x$ (which is equal to \eqref{discrim} by definition) 
 is given by 
\begin{equation}
 \Delta_2 = (2 \Lambda)^4 s_2^2 \{(s_1 + \Lambda)^2 - 4 s_2\} \{(s_1 - \Lambda)^2 - 4 s_2\}. 
\end{equation}
The factor $(2\Lambda)^4$ comes from that of resultant part \eqref{resultant}. 
The condition \eqref{n=2:s_2} corresponds to just the second order zero of $\Delta_2$ and we obtain 
\begin{equation}
  \left. \frac{\Delta_2}{(S_{22})^2}  \right|_{S_{22} = 0} = (2 \Lambda)^4 (s_1 + \Lambda )^2 (s_1 - \Lambda)^2, 
\end{equation}
 which again has the second order zero under the condition  \eqref{n=2:s_1}.  
Finally, we obtain 
\begin{equation}
 \left. \frac{\Delta_2}{S_{21\pm}^2 S_{22}^2}  \right|_{S_{21\pm} = S_{22} = 0} = (2 \Lambda)^6. 
\end{equation}

In the case of $n=3$, the curve is given by 
\begin{equation}\label{curve:n=3}
 y^2 = (\tilde{x}^3 + \tilde{s}_2 \tilde{x} + \tilde{s}_3)^2
  - \Lambda^2 \prod_{a=1,2} (\tilde{x} + \tilde{m}_a)^2 . 
\end{equation}
Putting $x = \tilde{x} + \tilde{m}_2$,  eq.\eqref{curve:n=3} becomes
\begin{equation}
 y^2 = (x^3 + s_1x^2 + s_2 x + s_3)^2 -\Lambda^2 x^2 (x+m)^2, 
\end{equation}
where $m = \tilde{m}_1-\tilde{m}_2$ and 
\begin{equation}
\begin{split}
 &s_1 = -3\tilde{m}_2 , \\ 
 &s_2 = 3\tilde{m}_2^2 + \tilde{s}_s, \\
 &s_3 = -\tilde{m}_2^3 - \tilde{s}_2 \tilde{m}_2 + \tilde{s}_3. 
\end{split}
\end{equation}
Let us set these parameters as follows: 
\begin{align}
 S_{31\pm} &:= s_1 \pm \Lambda = 0, \\ 
 S_{32\pm} &:= s_2 \pm \Lambda m = 0, \\
 S_{33}  &:= s_3 = 0. 
\end{align}
Then we obtain
\begin{equation} \label{curve:n=3c}
 y^2 = x^4 ( x^2 \mp 2 \Lambda x \mp 2\Lambda m ). 
\end{equation}
Hence, the curve \eqref{curve:n=3} is degenerate with multiplicity four.
Similarly for the $n=2$ case,  it can be confirmed that 
\begin{equation}\label{dis:n=3s}
 \left. \frac{\Delta_3}{S_{31\pm}^2 S_{32\pm}^2 S_{33}^2}  \right|_{S_{31\pm} = S_{32\pm} = S_{33} = 0} 
 =  (2 \Lambda m)^8 \tilde{\Delta}_{3\pm}, 
\end{equation}
where $\tilde{\Delta}_{3\pm} = 4\Lambda(\Lambda \pm 2m)$ is the discriminant of $x^2 \mp 2 \Lambda x \mp 2\Lambda m$. 
The condition $\tilde{\Delta}_3 = 0$ reduces \eqref{curve:n=3c} to 
\begin{equation}
 y^2 = x^4 ( x \mp \Lambda)^2. 
\end{equation}  

At general $n$, the curve is given by 
\begin{equation}
 y^2 = \left( \tilde{x}^{n} + \sum_{k=2}^n \tilde{s}_k \tilde{x}^{n-k}  \right) ^2
 - \Lambda^2  \prod_{a=1}^{n-1} (\tilde{x} + \tilde{m}_a)^2. 
\end{equation}
Putting $x = \tilde{x} + \tilde{m}_{n-1}$ and $m_a = \tilde{m}_a - \tilde{m}_{n-1}$, the curve becomes 
 \begin{equation}\label{curve:n}
  y^2 = \left( x^n + \sum_{k=1}^n s_k x^{n-k} \right)^2 
  - \Lambda^2 x^2 \left( \sum_{a=0}^{n-2} e_a x^{n-2-a} \right)^2, 
 \end{equation}
 where  
\begin{equation}
s_{k} = (-1)^k \mat{n\\k} \tilde{m}_{n-1}^k 
 + \sum_{j=2}^{k} (-1)^{k-j} \mat{n-j\\n-k} \tilde{s}_j \tilde{m}_{n-1}^{k-j}, ~~~~~
\end{equation}
and $e_a = e_a(m_1,m_2,\cdots,m_{n-2})$ is the $a$-th elementary symmetric polynomial.  
Let us demand the following conditions:   
\begin{align}\label{condition:n}
\begin{split}
 &S_{nk\pm} := s_k \pm \Lambda e_{k-1} = 0,~~~~k = 1,2,\cdots, n-1, \\
 &S_{nn} := s_n = 0. 
\end{split}
\end{align}
The curve \eqref{curve:n} becomes 
\begin{equation}
 y^2 = x^{n+1} \left\{ 
  x^{n-1} \pm 2\Lambda \sum_{a=0}^{n-2} e_a x^{n-2-a} 
 \right\}.
\end{equation}
It seems that the discriminant $\Delta_n $ for general $n$ satisfies 
\begin{equation}
\left. \frac{\Delta_n}{S_{nn}^2\prod_{k=1}^{n-1} S_{nk\pm}^2} \right|_{S_{n,k\pm}=S_{nn}=0} = 
 \left( 2\Lambda \prod_{a=1}^{n-2} m_a \right)^{2(n+1)} \tilde{\Delta}_{n\pm}, 
\end{equation} 
where $\tilde{\Delta}_{n\pm}$ is the discriminant of the remaining factor $y^2/x^{n+1}$. 
We have checked this relation up to $n=4$.

Hence, the S-W curve for $SU(n)$, $N_f=2n-2$ with the mass relations \eqref{mass} is maximally degenerate with multiplicity $n+1$ on the hypersurface satisfying \eqref{condition:n} in the parameter space. 
The conditions \eqref{condition:n} must be the very defining equations for the A-D critical hypersurface.

%%%%%%%%%%%%%%%%%%%%%%%%%%%%%%%%%%%%%%%%%%%%%%%%%%%% 
\section*{Acknowledgments}
%%%%%%%%%%%%%%%%%%%%%%%%%%%%%%%%%%%%%%%%%%%%%%%%%%%%
The work of HI is supported in part by JSPS KAKENHI Grant Number 19K03828 and by
the Osaka Metropolitan University (OMU) Strategic Research Grant 2022 (OMU-SRPP2022\_SU04).

%\appendix
%%%%%%%%%%%%%%%%%%%%%%%%%%%%%%%%%%%%%%%%%%%%%%%%%%%%
%\section{}
%%%%%%%%%%%%%%%%%%%%%%%%%%%%%%%%%%%%%%%%%%%%%%%%%%%%

%%%%%%%%%%%%%%%%%%%%%%%%%%%%%%%%%%%%%%%%%
%\bibliographystyle{arxiv}
%\bibliography{matrix}
%%%%%%%%%%%%%%%%%%%%%%%%%%%%%%%%%%%%%%%%%

%%%%%%%%%%%%%%%%%%%%%%%%%%%%%%%%%%%%%%%%%%%%%%%%%%%%

%%%%%%%%%%%%%%%%%%%%%%%%%%%%%%%%%%%%%%%%%%%%%%%%%%%%

%%%%%%%%%%%%%%%%%%%%%%%%%%%%%%%%%%%%%%%%%%%%%%%%%%%%
%%%%%%%%%%%%%%%%%%%%%%%%%%%%%%%%%%%%%%%%%%%%%%%%%%%%
\end{document}